\documentclass{iau307}
\usepackage{graphicx}
\usepackage{natbib}
\usepackage{url}
\usepackage{dtklogos}
\bibpunct{(}{)}{;}{a}{}{,}

\title[The Extremely Slow Rotation of $\xi^1$ CMa] 
{$\xi^1$ CMa: An Extremely Slowly Rotating Magnetic B0.7 IV Star}

\author[Shultz et al.]   
{Matt Shultz$^{1,2,3}$, Gregg Wade$^3$, Thomas Rivinius$^1$ \\ Wagner Marcolino$^4$, Huib Henrichs$^5$, Jason Grunhut$^1$ \\ \and the MiMeS Collaboration}

\affiliation{$^1$European Southern Observatory; $^2$Queen's University, Canada \\[\affilskip $^3$Royal Military College, Canada  ]
 $^4$Observat\'orio do Valongo, UFRJ, Brazil; $^5$Anton Pannekoek Institute for Astronomy, University of Amsterdam, the Netherlands \\ email: {\tt mshultz@eso.org}}

\pubyear{2014}
\volume{307} 
\pagerange{}
\jname{New windows on massive stars: asteroseismology, interferometry, and spectropolarimetry}
\editors{G. Meynet, C. Georgy, J.H. Groh \& Ph. Stee, eds.}

\begin{document}

\maketitle

\begin{abstract}
We present our analysis of 6 years of ESPaDOnS spectropolarimetry of the magnetic $\beta$ Cep star $\xi^1$ CMa (B1 III). This high-precision magnetometry is consistent with a rotational period $P_{\rm rot} > 40$ yr. Absorption line profiles can be reproduced with a non-rotating model. We constrain $R_*$, $L_*$, and the stellar age via a Baade-Wesselink analysis. Spindown due to angular momentum loss via the magnetosphere predicts an extremely long rotational period if the magnetic dipole $B_{\rm d} > 6$ kG, a strength also inferred by the best-fit sinusoids to the longitudinal magnetic field measurements $B_{\rm Z}$ when phased with a 60-year $P_{\rm rot}$. 
\end{abstract}

\firstsection 
\section{Magnetometry}

The B0.7 IV $\beta$ Cep star $\xi^1$ CMa was detected as magnetic by FORS1 \citep{hubrig2006} and later confirmed by ESPaDOnS \citep{silvester2009}. Further FORS1/2 observations yielded a 2.2 d period \citep{hub2011a}, however many stars for which periods were reported based on FORS1/2 data were not confirmed to be magnetic by ESPaDOnS data \citep{shultz2012}, casting doubt on the reliability of period analysis using FORS1/2 data. An analysis of an earlier, smaller ESPaDOnS data-set was presented by \cite{fr2011}, who found $P_{\rm rot} \sim$4 d. 

\begin{figure}[h]
\begin{center}
\includegraphics[width=1.\textwidth]{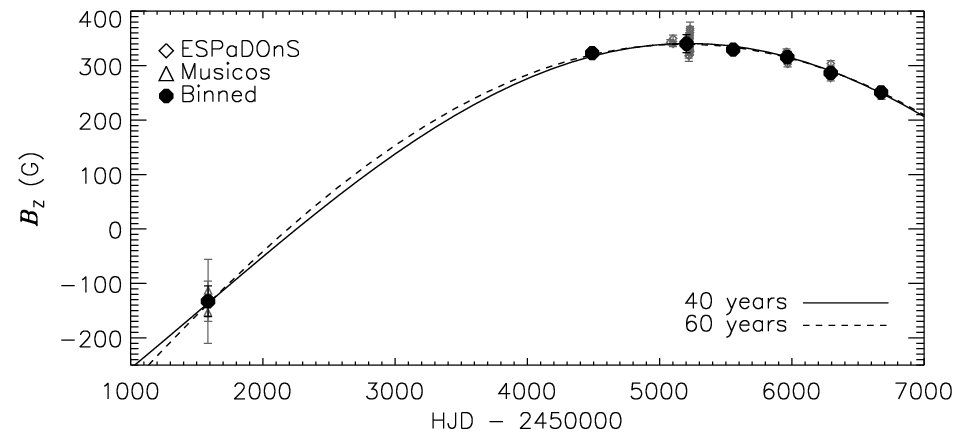} 
\caption{Longitudinal magnetic field $B_{\rm Z}$ measurements as a function of time. Measurements binned by epoch are in solid black circles. }
\label{fig1}
\end{center}
\end{figure}

The ESPaDOnS dataset has grown to 34 Stokes $V$ spectra obtained from 2008 to 2014, with annual clusters of up to 16 spectra separated by hours to days, and uniform integration times of 240 s. Sharp spectral lines and high SNR ($>$440) lead to a mean error bar in $B_{\rm Z}$ of $<\sigma_B> = 6$ G in the single-spectrum least-squares deconvolution (LSD) profiles \citep{koch2010}. $B_{\rm Z}$ declines smoothly from 340$\pm$17 G in 2009 to 251$\pm$3 G in 2014 (see Fig. \ref{fig1}). There is a systematic error ($\sim$15 G) due to a small variation of $B_{\rm Z}$ with the pulsation period. We have also located 2 archival MuSiCoS observations obtained in 2000. In both spectra, the polarity of the Zeeman signature is negative, with $B_{\rm Z} = -137 \pm 32$ G. {\em The minimum $P_{\rm rot}$ compatible with the combined datasets is 40 years: the longest rotation period yet found for an early B-type star}.

\section{Analysis}

We used the Si {\sc iii} 455.3 nm line to derive $v\sin{i}$ and macroturbulence $\zeta$, performing a goodness of fit test using a grid of synthetic line profiles including radial pulsation \citep{saesen2006}, finding $v\sin{i} < 6$ km/s, and $\zeta = 20 \pm 3$ km/s, i.e. the line profile can be be reproduced with a non-rotating model.


To constrain the stellar parameters we performed a Baade-Wesselink analysis, combining the pulsation phase-dependant variations in integrated light (via Hipparcos photometry), radial velocity (via ESPaDonS and CORALIE spectra, \citealt{saesen2006}), and $T_{\rm eff}$ (via EW ratios for various elemental ionic species). $<T_{\rm eff}> = 25.9\pm 0.1$ kK, with a $\pm 0.5$ kK amplitude variation. With the bolometric correction of \cite{nieva2013}, the model light curve reproduces the photometric variability with $R = 8.7\pm0.7 R_\odot$, leading to $\log{L/L_\odot} = 4.46\pm0.07$, where the uncertainty is largely a function of the error in the Hipparcos parallax. Isochrones \citep{ekstrom2012} indicate an age of 12.6 Myr.


Magnetic stars shed angular momentum via the torque applied to the stellar surface by the corotating magnetosphere \citep{wd1967, ud2009}. Can a magnetic early B-type star spin down to $P_{\rm rot} > 40$ yrs after only 12.6 Myr? We computed the Alfv\'en radius \citep{ud2002} based on the mass-loss rate and wind terminal velocity predicted by the recipe of \cite{vink2001} using the stellar parameters above, for values of $B_{\rm d}$ between 1.2 and 8.0 kG, and from this the spindown time \citep{ud2009}, assuming initially critical rotation. If $B_{\rm d} > 6$ kG, $P_{\rm rot} >$ 40 yrs can be achieved; if $P_{\rm rot} = 60$ yrs, the best-fit sinusoid to $B_{\rm Z}$ implies $B_{\rm d} \sim 6$ kG.

\bibliographystyle{iau307}
\bibliography{bib_dat.bib}







\end{document}